\documentclass[pre,groupaddress,twocolumn,showpacs,floatfix]{revtex4}
\usepackage{epsfig,amsmath,amssymb,graphics,color,calc,bm}

\newcommand{\be}{\begin{equation}}
\newcommand{\ee}{\end{equation}}
\newcommand{\ba}{\begin{eqnarray}}
\newcommand{\ea}{\end{eqnarray}}

\newcommand{\Diso}{\Delta^\mathrm{iso}}
\newcommand{\Dprop}{\delta}
\newcommand{\Cg}{C_\mathrm{g}}

\newcommand{\Eth}{\mathbb{E}}

\newcommand{\notsigma}{\ell}
\newcommand{\riso}{\rangle_\mathrm{iso}}

\begin{document}

\title{Structure and dynamics in glass-formers: predictability
at large length scales}

\author{Ludovic Berthier}
\affiliation{Joint Theory Institute, Argonne National Laboratory and
University of Chicago,
5640 S. Ellis Av., Chicago, Il 60637}

\altaffiliation{Permanent address: 
Laboratoire des Collo\"{\i}des, Verres et Nanomat\'eriaux, 
UMR 5587, Universit{\'e} Montpellier II and CNRS, 34095 Montpellier, France}

\author{Robert L. Jack}
\affiliation{Department of Chemistry, 
University of California, Berkeley, CA 94720-1460}

\date{\today}

\begin{abstract}
Dynamic heterogeneity in glass-formers
has been related to their static structure using the 
concept of dynamic propensity. 
We re-examine this relationship by analyzing dynamical  
fluctuations in two atomistic glass-formers and two
theoretical models. 
We introduce quantitative statistical indicators
which show that the dynamics of individual particles cannot be predicted
on the basis of the propensity, nor by any structural indicator.
However, the spatial structure of the 
propensity field does have predictive power for the spatial correlations
associated with dynamic heterogeneity. 
Our results suggest that the quest for a 
connection between static and dynamic properties 
of glass-formers at the particle level is vain, but they 
demonstrate that such connection does exist on larger length scales.
\end{abstract}


\maketitle

{\it The future ain't what it used to be}---Y. Berra 

\section{Introduction}
\label{introduction}

Supercooled liquids near the glass transition have paradoxical physical
properties~\cite{SRN,DH}. 
They exhibit a range of peculiar dynamical features that 
have been attributed to spatially heterogeneous 
dynamical relaxation~\cite{DH,Kob1,Ediger_blocks,GC,chi4_franz,heuer,chi4_ton,
Berthier_science_long,Whitelam04,pinaki}, but their structure, as measured by 
two-point correlation functions, appears homogeneous and unspectacular.  
Theoretical pictures of the glass transition assume different
kinds of connections between static and dynamic properties.
For example, in a picture based on 
dynamical facilitation~\cite{GC,FA84},
one postulates the existence of mobile and immobile 
regions, with the implicit assumption that these regions
have a structural origin. 
Frustration-based theories~\cite{frustration_review} 
infer dynamical behavior by assuming the existence of domains with
a preferred local order.
Alternatively, one can attempt to
connect static and dynamical properties through the configurational 
entropy~\cite{AG,KTW}, through two-point density 
correlations~\cite{MCT}, through elastic properties~\cite{modes,reichman},
or through the idea of a rough energy 
landscape~\cite{heuer,Stillinger-Weber}.  
The extent to which these connections can be 
objectively established in experiments 
and computer simulations is an important criterion for evaluating
different theoretical pictures.

To address this point,
Harrowell and co-workers introduced the 
isoconfigurational ensemble~\cite{Harrowell}, which
isolates the effect of the liquid structure on its
dynamical fluctuations. 
In this statistical ensemble, dynamical observables known as `propensities' 
are obtained by averaging over
independent trajectories from the same initial configuration.  
Fixing the initial particle positions preserves
structural information on all length scales, 
and allows the influence of structure to be separated
from intrinsically dynamical fluctuations.
The difficult task of connecting structure to dynamics is then broken
into two seemingly simpler ones: first connect structure to propensity, 
then propensity to dynamics.  

This idea has been exploited in several works~\cite{Harrowell,Harro07,
frechero,kennet,hedges,coslovich,appig,poole}, which focussed
primarily on the first stage of the problem, and aimed at 
finding correlations between propensity and local or non-local 
structural quantities.
In this paper, we concentrate instead on the second stage, the 
connection between propensity and dynamics, which has 
received relatively little attention.
After all, if propensity and dynamics only 
had weak connections, the isoconfigurational ensemble would not be 
such a useful tool.
The isoconfigurational average preserves all fluctuations that 
have any connection to the initial structure, but fluctuations whose
origin is inherently dynamical are averaged away.  
Thus, propensities capture the physically relevant dynamical 
fluctuations if and only if these fluctuations are 
structural in origin~\cite{Harrowell}, in which case
dynamics can be predicted from knowledge of the structure. 
In concentrating on the connection between structure and propensity, 
previous works~\cite{poole,appig,coslovich,hedges,frechero,kennet} 
have assumed that this is indeed the case. 

Here, we analyze quantitatively the implicit assumption 
of predictability.
We establish that the dominant single-particle dynamical
fluctuations are intrinsically dynamical, and
not linked with liquid structure. 
However, we find that the sizes  
and shapes of mobile and immobile regions can be predicted 
from the propensity, and these collective dynamical
fluctuations do indeed have a structural origin. 
Although the relevance of intrinsically dynamical fluctuations
has been discussed in a qualitative way 
by Widmer-Cooper and Harrowell~\cite{Harrowell,Harro07}, 
the subtle length scale dependence of the correlation between 
structure and dynamics, and therefore of predictability, 
was not discussed nor anticipated in previous work.  

After defining our models in Sec.~\ref{models},
we discuss single particle and collective dynamics 
in Secs.~\ref{single} and \ref{collective}, combining
illustrative snapshots with quantitative analysis,
and comparing our results with the behavior
in schematic models.  We conclude
in Sec.~\ref{conclusion}, and identify 
directions for future study.

\section{Models}
\label{models}

We present numerical data for two atomistic glass-formers:
a Lennard-Jones (LJ) binary mixture~\cite{KA},
and a model of silica due to Beest, Kramer
and van Santen (BKS)~\cite{BKS}.  We use Monte Carlo 
dynamics, which have been shown to yield 
a dynamical behavior in excellent agreement with Newtonian 
dynamics when time is scaled 
appropriately~\cite{Berthier_MC_LJ,Berthier_MC_BKS}.
We measure time in Monte Carlo sweeps, 
and other units are as in Refs.~\cite{Berthier_MC_LJ,Berthier_MC_BKS}.
The LJ system has 1000 particles, and we study temperatures
between $1.0$, where the system is a simple liquid, and $0.47$,
where the relaxation time has increased by a factor of approximately
$1000$, and the system is in the glassy regime. For reference
the mode-coupling temperature for this system is $T_\mathrm{c}=0.435$.
The BKS system has 1008 atoms, and the relaxation time also
spans around three decades, from a liquid state
around 6000~K to a glassy one around 3000~K.

We also study the one-spin facilitated
Fredrickson-Andersen (1-FA) model~\cite{FA84}, using Monte Carlo
simulations.  It represents a simple model of a dynamically
heterogeneous material, with a few mobile regions that facilitate
motion in immobile regions nearby~\cite{GC}.  Defining
spins $n_i\in\{0,1\}$ on the site of a lattice, we identify
sites with $n_i=1$ as `mobile' and those with $n_i=0$ as
`immobile'.  The dynamics of the system obey detailed balance with 
respect to a trivial energy function $E=\sum_i n_i$.  The non-trivial
behavior of the model comes from a dynamical constraint:
spins are allowed to flip only if at least one of their
neighboring sites is mobile.  At low temperatures, $T<1$,
and low spatial dimension, $d<2$,
the model displays glassy features such as  
transport decoupling~\cite{Jung04} 
and an increasing dynamical length scale~\cite{Whitelam04}.

We use $C_i(t)$ to denote a general
dynamical object attached 
to particle $i$, such as $f_i(t) \equiv  
\cos ( \bm{k} \cdot [ {\bm{r}}_i(t) - {\bm{r}}_i(0)] )$ 
or
$\mu_i(t) \equiv | {\bm{r}_i(t)} - {\bm{r}_i(0)} |$.
For $f_i(t)$, we use $|\bm{k}|=6.7$ for the LJ
model and $|\bm{k}|=1.8$~\AA$^{-1}$ for silica.
These wave vectors correspond to the location of the first 
diffraction peak in the LJ system, and to the pre-peak 
of the structure factor of silica.  The quantity $f_i(t)$
is therefore a measure of local relaxation in these liquids.
In lattice models, we consider $P_i(t)$, the persistence function on 
site $i$, which takes the value of unity 
if spin $i$ has not flipped in the interval
$[0, t]$, and zero otherwise. 

We define an isoconfigurational average
$ \langle \cdots \riso$ in which the initial positions
of all particles are held fixed~\cite{Harrowell}, but dynamical trajectories
are made independent through the use of different random 
numbers in the Monte Carlo trajectories~\cite{hedges}.
The dynamic propensity is then 
defined by $\langle C_i(t) \riso$~\cite{Harrowell}.
Equilibrium ensemble averages are denoted by $\Eth[\cdots]$. As usual, 
we define the structural relaxation
time $\tau_\alpha$ by $\Eth[f_i(\tau_\alpha)]=1/{\rm e}$.

\section{Single particle dynamics}
\label{single}

\subsection{Atomistic systems}

\begin{figure}[b]
\epsfig{file=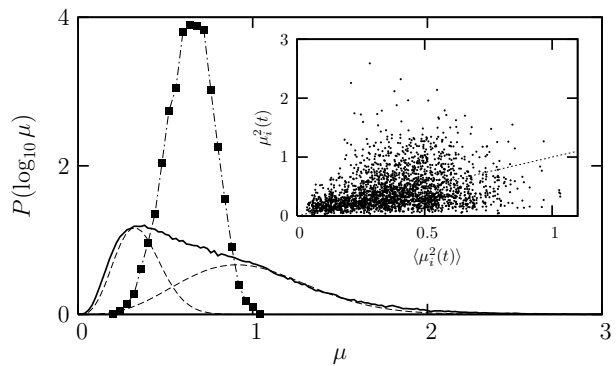,width=8.cm}
\caption{\label{scatter} Data for
the LJ system in the glassy regime ($T=0.47$).  
Inset: Scatter plot, comparing
squared displacements in a single run, $\mu_i^2(\tau_\alpha)$, 
with their average over runs from the same initial 
condition, $\langle \mu_i^2(\tau_\alpha) \riso$. 
Note that the lack of correlation between propensity and dynamics
necessitates the 
use of different scales on the two axes.
The dotted line is $\mu_i^2(\tau_\alpha)=
\langle\mu_i^2(\tau_\alpha)\riso$.
Main:~The broad distribution of particle (logarithmic) displacements,
$P(\log \mu)$ (full line) arises
from the coexistence of mobile and immobile particles,  
illustrated with dashed lines (Gaussian distributions).
This is compared with the much narrower distribution of propensities, 
$P(\log \langle \mu \riso)$.} 
\end{figure}

We begin with a qualitative analysis of fluctuations within
the isoconfigurational ensemble.  Fig.~\ref{scatter} 
shows a scatter plot for the LJ system, comparing
squared particle displacements in a single run, $\mu_i^2(t)$, 
with their corresponding propensities, $\langle \mu_i^2(t) \riso$. 
Points near the dashed line represent particles whose
displacements in this single run are close to their 
isoconfigurational averages.
Given the large scatter seen in Fig.~\ref{scatter}, it is 
clear that the propensity cannot be used to predict the 
actual value of the displacement in a single run. That is,
even when all aspects of the initial structure are held constant,
there are large fluctuations in
the single-particle dynamics, which appear to be more 
important than particle-to-particle fluctuations of the propensity.
These large dynamical fluctuations mean that 
the propensities have very little predictive power 
for the single-particle dynamics.

Fig.~\ref{scatter} also shows
the distribution of particle (logarithmic) displacements, $P(\log \mu)$, 
and of propensities, $P(\log \langle \mu \riso)$.
The former, which is directly related to the van-Hove function,
has a broad, non-Gaussian shape, reflecting the coexistence
of mobile and immobile particles in the liquid~\cite{Kob1,pinaki},
as illustrated by fitting the large-$\mu$ and small-$\mu$
parts of the distribution with two distinct Gaussian distributions.
The latter distribution (particle propensities)
is much narrower and structureless. In particular, 
the distinction between fast and slow particles is no longer apparent.  
This most distinctive feature of dynamic heterogeneity~\cite{DH} 
is therefore not structural in nature, but is intrinsically dynamical.  

The details of the distribution of the propensity shown 
in Fig.~\ref{scatter} are different from those shown in~\cite{Harrowell,
Harro07}.  We attribute this to differences between
our model systems (in particular their different 
dimensionalities and tendencies to crystallize),
and to our use of slightly different observables (we use
simple distances while squared distances 
were used in~\cite{Harrowell,Harro07}, emphasizing particles
with large displacement).  For our purposes,
the important feature is that the distribution of
propensities is much narrower than that of bare displacements, 
which is consistent with earlier results~\cite{potato}.

We now support these qualitative statements by quantitative measures.
In order to disentangle structural and dynamical sources
of fluctuations, we define three variances:
\ba
\Dprop_C(t) &=& \Eth\left[ \langle C_i(t) \riso^2 \right] - \Eth^2[C_i(t)],  
\nonumber \\ 
\Diso_C(t) & = & \Eth\left[ \langle C_i^2(t) \riso-
  \langle C_i(t) \riso^2 \right] \nonumber, \\ 
\Delta_C(t) & = & \Eth\left[ \langle C_i^2(t) \riso\right] -
 \Eth^2\left[ C_i(t) \right], 
\label{def1}
\ea
so that $\Delta_C(t)=\Diso_C(t) + \delta_C(t)$; we use the short-hand
notation $\Eth^2[ \cdots] = ( \Eth [ \cdots ] )^2$ and
the subscript indicates the dynamic
observable of interest. Thus, 
$\Dprop_C(t)$ measures particle to particle 
fluctuations of the propensity~\cite{appig} and captures the 
structural component of the fluctuations. 
$\Diso_C(t)$ measures the fluctuations 
of $C_i(t)$ between different runs at fixed 
initial configuration and captures therefore the dynamical 
component of the fluctuations.
Their sum $\Delta_C(t)$ naturally measures the total 
amount of fluctuations.

These quantities are defined in the same spirit as the ensemble-dependent 
susceptibilities of Refs.~\cite{Berthier_science_long}.   
To assess the influence of a given variable (the structure) 
on the fluctuations measured by $\Delta_C(t)$, 
we use a constrained statistical ensemble (the isoconfigurational
ensemble) where this variable is kept fixed, and measure
a `restricted' variance, $\Diso_C(t)$. The difference
between restricted and unrestricted variances
is $\delta_C(t)$, which accounts for the fluctuations
of the restricted
variable~\cite{Berthier_science_long}. 
In our case, the relative sizes of $\delta_C(t)$ and $\Diso_C(t)$
quantify the relative influence of structural and dynamical
fluctuations.
It is natural to introduce the dimensionless ratio
\be
R_C(t) = \frac{\delta_C(t)}{\Delta_C(t)},
\ee
with $0 \leq R_C(t) \leq 1$. Small values
of $R_C(t)$ mean that the structure has little effect on the
single particle dynamics, large values mean instead that 
structure is a very good predictor of the dynamics.

\begin{figure}
\psfig{file=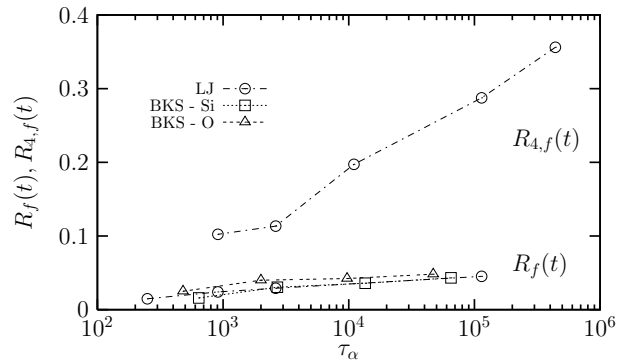,width=8cm}
\caption{\label{ratiolj} Ratios $R_f(\tau_\alpha)$ 
for the LJ and silica systems, 
and $R_{4,f}(\tau_\alpha)$ for the LJ system.  Temperatures
are  
$0.47\leq T\leq1.0$ (LJ) and $3000\mathrm{K}\leq T\leq6100\mathrm{K}$ (BKS).
$R_f(t)$ is small and grows slowly with decreasing
$T$, while $R_{4,f}(t)$ is large and grows steadily.  This 
suggests that connections between structure 
and dynamics are only significant on large length scales.}
\end{figure}

We present results for the ratio $R_f(\tau_\alpha)$ 
for LJ and BKS glass-formers
in Fig.~\ref{ratiolj}.  This ratio vanishes
at high $T$, and grows slowly as $\tau_\alpha$ 
increases, reaching about 4\% in both systems 
at the lowest temperature studied (when $\tau_\alpha$ has grown by 
about 3 decades). The smallness
of $R_f(\tau_\alpha)$
confirms the impression gained from
Fig.~\ref{scatter}: structure at time 0 has little influence on the dynamics
of individual particles at times $\tau_\alpha$. 

\subsection{Comparison with a schematic model}
\label{subsec:schem}

It is instructive to evaluate $R_C(t)$ in a simple
model of a dynamically heterogeneous liquid.  
We suppose that particles
diffuse independently with diffusion constants $D_i$
that depend on their initial
environments. We also assume that
effects
of the initial environment decay on timescales comparable
with $\tau_\alpha$. Then, each particle has
$\langle f_i(t) \riso=e^{-D_i k^2 t}$, and
\begin{equation}
F(t)\equiv\Eth[ f_i(t) ]
= \int_0^\infty \! \mathrm{d}\lambda\, g(\lambda) \exp(-\lambda t),
\label{dis}
\end{equation}
where $\lambda_i=D_i k^2$ is a rescaled diffusion
constant distributed according to 
$g(\lambda)$.   
Using the definition
of $f_i(t)$, we have $f_i(t)^2=(1/2)\{\cos(2\bm{k}\cdot[\bm{r}_i(t)-
\bm{r}_i(0)])-1\}$, and we find
\ba
\Dprop_f(t)&=&F(2t)-F^2(t), \nonumber\\
\Delta_f(t)&=&(1/2)[1 + F(4t) - 2F^2(t)].
\ea

In a dynamically homogeneous system,
all particles have the same diffusion constant, $F(t)$
decays exponentially, and $\Dprop_f(t)=0=R_f(t)$.
On the other hand, consider a heterogeneous system
with equal populations of fast and slow particles,
whose rescaled diffusion constants are $\lambda_1$ and $\lambda_2$.
Using Eq.~(\ref{dis}), this leads to a two-step decay: $F(t)=
(1/2)[e^{-\lambda_1 t}+e^{-\lambda_2 t}]$.   
In this case, $R_f(t)$ has a non-monotonic 
time dependence, vanishing at small and long times, 
with a maximal value near $50\%$
during the plateau of $F(t)$ (that is, for times such that 
$\lambda_1^{-1} \ll t 
\ll \lambda_2^{-1}$).
This indicates that correlations 
between structure and dynamics are generically maximal during
plateaux of $F(t)$~\cite{appig}; our atomistic simulations
are also consistent with this indication. 

Our analysis of this highly schematic model
demonstrates that large values of $R_f(t)$ are
obtained in the presence of reproducibly fast and slow particles. 
Comparing this reference theory with the small values
for $R_f(t)$ shown 
in Fig.~\ref{ratiolj}, it follows that
the correlation between structure and single-particle dynamics is 
weak in atomistic systems and does not seem to dramatically
increase when temperature gets smaller.

We conclude that the search for a connection between static and dynamic
properties at the single-particle level is in vain.  Certainly,
no such connection has been 
found~\cite{Harrowell,Harro07,frechero,hedges,coslovich,kennet,appig,poole}.

\section{Collective dynamics}
\label{collective}

Having ruled out the structural origin of one important aspect
of dynamic heterogeneity, we now address a different question:
Do spatial fluctuations of the propensity carry meaningful 
information on the geometry and spatial extent of the dynamic
heterogeneities in supercooled liquids? We will show that
while it is not possible to use the structure to predict 
whether a given particle will be fast or slow in a single run, 
it is possible to tell if it belongs to a fast or slow 
{\it region}.

\subsection{LJ system}

We begin by generalizing the variances in (\ref{def1}) to those of 
global dynamic quantities, $\Cg(t) = N^{-1} \sum_i C_i(t)$. We define:
\ba
\Dprop_{4,C}(t) & = & N \{  \Eth\left[ \langle \Cg(t) \riso^2 \right] -
 \Eth^2\left[ \Cg(t) \right] \} ,  \nonumber \\ 
\chi^\mathrm{iso}_{4,C}(t) & = & N \{  \Eth\left[ \langle \Cg^2(t) \riso- 
\langle \Cg(t) \riso^2 \right]\}, \nonumber \\ 
\chi_{4,C}(t) & = & N \{ \Eth\left[ \langle \Cg^2(t) \riso \right] -
 \Eth^2\left[ \Cg(t) \right] \} ,
 \label{def2}
\ea
with $\chi_{4,C}(t)=\delta_{4,C}(t)+\chi^\mathrm{iso}_{4,C}(t)$. 
The usual four-point susceptibility $\chi_{4,C}(t)$
measures the size of collective dynamical
fluctuations~\cite{chi4_franz,chi4_ton}.  By analogy
with $\Dprop_C(t)$ and $\Diso_C(t)$, we identify
$\Dprop_{4,C}(t)$ as the contribution to $\chi_{4,C}(t)$ associated
with the structure, and $\chi^\mathrm{iso}_{4,C}(t)$ as the intrinsically
dynamical contribution.
Then, in analogy with $R_C(t)$,
\be
R_{4,C}(t) = \frac{\delta_{4,C}(t)}{\chi_{4,C}(t)},
\ee
is a dimensionless
measure of the effect of the structure on collective
aspects of the dynamics.
In Fig.~\ref{ratiolj}, we show that the contribution 
of $\delta_{4,f}(\tau_\alpha)$ to the dynamic fluctuations is 
about 35\% at $T=0.47$ (to be compared to the 4\% found for single particle 
heterogeneity).  Moreover, $R_{4,f}(\tau_\alpha)$ 
grows steadily when $T$ decreases.
This quantitative measurement
shows that when the system is in the glassy regime,
collective dynamical fluctuations
are indeed quite reproducible in repeated runs from the same initial
configuration.  We conclude
that connections between structural and dynamical 
properties are significant on these length scales,
and that the connection even gets stronger as the relaxation
time increases.

To investigate the fluctuations associated with this effect,
we interpolate between the single-particle and collective
dynamics, by averaging over a length scale $\notsigma$:
\begin{equation}
\overline{C}_i(t,\notsigma) = \frac{\sum_j C_j(t) h( | \bm{r}_i-\bm{r}_j |  )}{
                     \sum_j h( | \bm{r}_i-\bm{r}_j |  )},
\quad h(x) = e^{-(x/\notsigma)^2}.
\end{equation}
By coarse-graining in this way, we can investigate how the reproducibility
of dynamical fluctuations varies with length scale.

\begin{figure}
\psfig{file=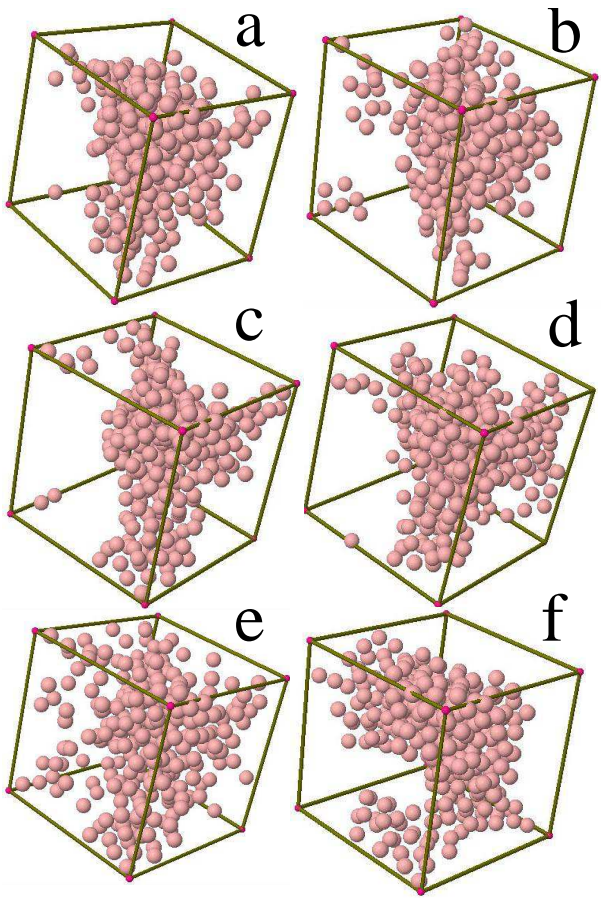,width=8.cm}
\caption{\label{4fig} 
a:~Image of the $300$ particles
with the smallest propensity for motion (largest values
of $\langle f_i(\tau_\alpha) \riso$),  
for a representative initial condition in the 
LJ system in the glassy regime ($T=0.47$).  
b-d:~Images of the $300$ particles with
largest values of the coarse-grained observable
$\bar{f}_i(\tau_\alpha,\notsigma=1)$,
in three representative runs from the same initial
condition.  Spatial structure similar to that of
(a) is found in all cases. 
e:~As (c), without coarse-graining ($\notsigma=0$).
f:~Image of the $300$ particles with
the smallest values of the coarse-grained potential energy, 
$\overline{e}_i(\notsigma=2)$, for the same configuration.
The resulting
structure is correlated with that of (a).}
\end{figure}

In Fig.~\ref{4fig}, we use snapshots of the system
to illustrate that large scale features of the dynamics
are indeed reproducible from run to run.  For a representative initial 
configuration at a low temperature, Fig.~\ref{4fig}a shows the 
LJ particles with the largest values of the propensity 
$\langle f_i(\tau_\alpha)\riso$
(these are the particles which are slow on average).  
In all panels of Fig.~\ref{4fig},
we show about 1/3 of the particles:  
this threshold is small enough to
give clear images, but large enough to avoid
placing undue emphasis on rare fluctuations. 

Figs.~\ref{4fig}b-d show particles 
with the largest values of $\overline{f}_i(t,\notsigma=1)$
in three individual runs.
Loosely speaking, the coarse-graining scale $\notsigma=1$ means that
$\overline{f}_i$ measures how much motion is associated
with a particle and its nearest neighbors, on the
time scale $t$ (the choice of this length scale is discussed 
below).  Thus, Figs.~\ref{4fig}b-d show particles
that are located in a relatively immobile environment during these
three trajectories.  Similar clusters of immobile particles
are apparent in all three trajectories, and these
clusters correlate quite well with the cluster of
slow particles that is observed in the propensity map
in Fig.~\ref{4fig}a.  Since the slow behavior
of these clusters is reproducible in independent runs
from the same initial condition, it surely must have
a structural origin.  This is consistent with the rather
strong coupling of structure and collective dynamics that
was identified in Fig.~\ref{ratiolj}.
We also note in passing
that the coarse-grained pictures in Figs.~\ref{4fig}b-d
are computationally much cheaper than calculating the
propensity field.

\begin{figure}
\psfig{file=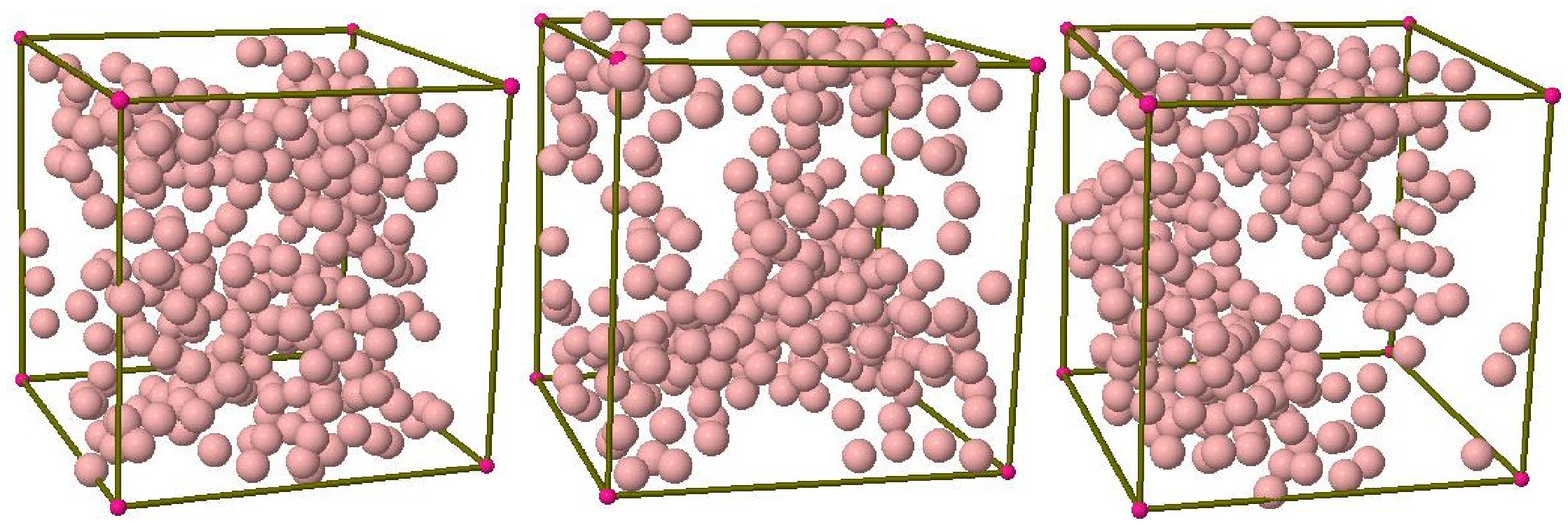,width=8cm}
\psfig{file=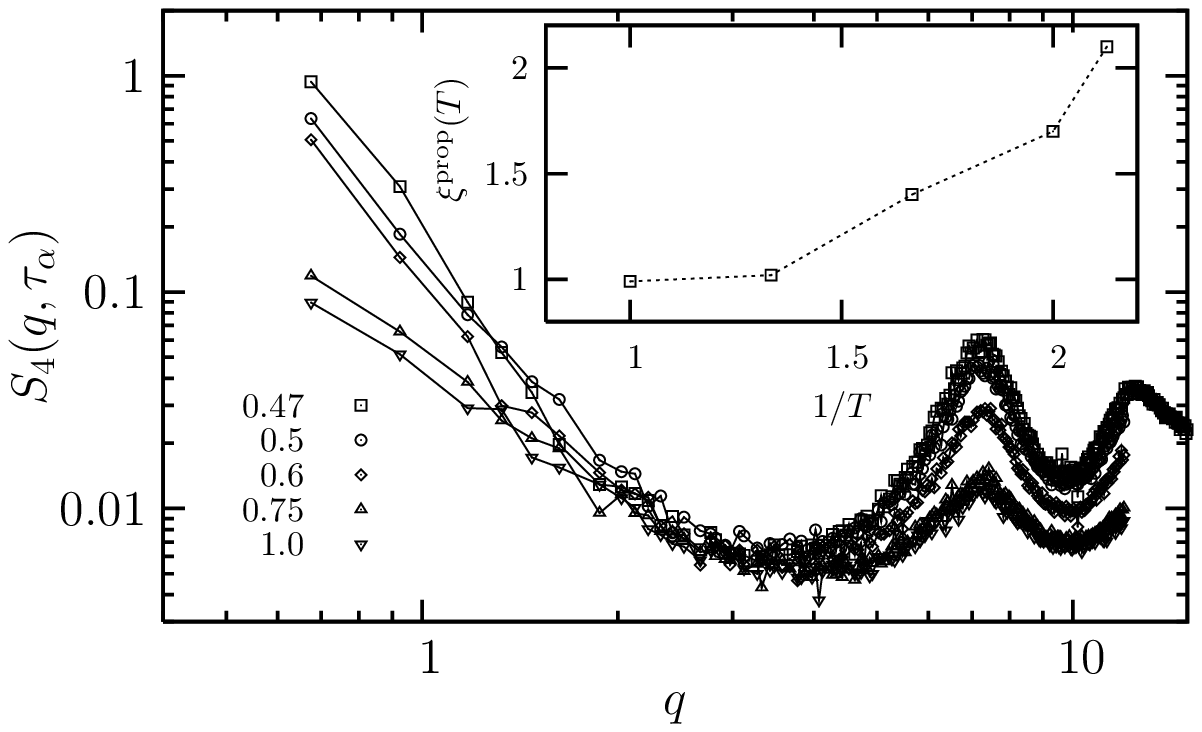,width=7.5cm}
\caption{\label{s4} Top: Images of the $300$ particles 
with the largest values of
$\langle f_i(\tau_\alpha)\riso$
at $T=1.0$, 0.6, and 0.47 (left to right) in the LJ system.
Increasing clustering of propensity fluctuations 
is evident.
Bottom: Structure factor of propensity fluctuations
at several $T$. The inset shows the extracted lengthscale
$\xi^\mathrm{prop}(T)$, 
which grows as the temperature decreases.}
\end{figure}

As discussed above, the coupling between structure
and dynamics at the single-particle level is weak.  This
is further illustrated in Fig.~\ref{4fig}e, where we have not
coarse-grained, but simply identified slow particles
by their values of $f_i(\tau_\alpha)$, again using
the same initial condition.  In this case 
the immobile cluster that is apparent in  Fig.~\ref{4fig}a-d
is obscured by large intrinsically dynamical fluctuations.
Comparing Fig.~\ref{4fig}e with Figs.~\ref{4fig}b-d
shows that the effect of coarse-graining on the
short length scale $\notsigma=1$ is quite
effective in suppressing these fluctuations, allowing
the slow cluster to become apparent.  When
coarse-graining in this way, we must also ensure that 
the length scale $\notsigma$ is smaller than the correlation
length associated with the dynamically correlated clusters, or
else these clusters will themselves be obscured.  
We establish below (Fig.~\ref{s4}) that the
dynamically correlated clusters have a length scale 
$\xi^\mathrm{prop}\simeq2$ at this temperature.  
Thus, while it would be desirable
to have well-separated length scales
$\notsigma$ and $\xi^\mathrm{prop}$, we can at least
establish that $\notsigma<\xi^\mathrm{prop}$, as required
for the consistency of our analysis.  As expected,
we find that on further increasing the coarse-graining
scale $\notsigma$, the structure of the immobile clusters
in Figs.~\ref{4fig}b-d is still apparent, but the ability to
resolve their shape is lost.  

Returning to the spatial correlations of the propensity,
we interpret $\delta_{4,f}(t)$ as a
dynamic susceptibility associated with spatial fluctuations 
of the propensity, by analogy
with the four-point susceptibility 
$\chi_{4,f}(t)$~\cite{chi4_franz,chi4_ton}. That is,
defining the fluctuations of the propensity by
$\langle \delta f_i(t)\riso = \langle f_i(t)\riso- F(t)$, then
the spatial correlation function of the propensity is
\begin{eqnarray}
G_4^\mathrm{prop}(\bm{r},t) &=& 
\Eth\Big[
  N^{-1} \sum_{ij} 
 \langle \delta f_i(t)\riso \langle \delta f_j(t)\riso \
 \nonumber\\
& & \qquad \qquad \times
  \delta(\bm{r}_i(0) - \bm{r}_j(0)-\bm{r})
\Big].
\end{eqnarray}
The associated structure factor is the Fourier transform
of this function:
\begin{eqnarray}
S_4^\mathrm{prop}({\bf q},t) 
&=& \int\!\mathrm{d}\bm{r}\, 
e^{i\bm{q}\cdot\bm{r}}
G_4^\mathrm{prop}(\bm{r},t),
\end{eqnarray}
and the associated dynamical susceptibility is
$\delta_{4,C}(t)=S_4^{\rm prop}(q \to 0,t)$.  
By analogy with the four-point susceptibility,
we expect $\delta_{4,C}(t)$ to be proportional to
the number of particles associated with collective
fluctuations of the propensity.

Like $\chi_4(t)$, the susceptibility of the propensity $\delta_{4,C}(t)$
is a non-monotonic function of time that peaks near $\tau_\alpha$.
We find that the height of this peak
grows as $T$ decreases,
suggesting increasing spatial correlations of the propensity.
This is illustrated by the images in Fig.~\ref{s4}, which show that 
the particles with smallest propensities are increasingly spatially 
clustered as $T$ decreases. To confirm this visual impression, 
we present in Fig.~\ref{s4} our numerical 
data for $S_4^{\rm prop}(q,t)$ measured
at different temperatures in the LJ system. The structure factor of 
the propensity is similar to the structure factor 
of dynamic heterogeneity, with the striking appearance of 
a small-$q$ peak. Following earlier work we  
estimated~\cite{foot} the correlation
length of propensity fluctuations $\xi^{\rm prop}(T)$, 
as shown in the inset of Fig.~\ref{s4}. It has
a clear, but rather slow, growth when $T$ decreases, compatible 
with determinations of a typical lengthscale
of dynamic heterogeneity~\cite{Berthier_science_long,GC,chi4_ton}.

These measurements confirm that spatial correlations of the propensity 
increase when temperature decreases, just as the length scale  
of dynamic heterogeneity does. Therefore, we find the intriguing 
result that the value of the propensity of any individual 
particle is only weakly correlated to its dynamical behavior, but 
the spatial correlations of these propensities do carry information
about spatially heterogeneous dynamics.  In short, the
spatial structure of the propensity maps in Ref.~\cite{Harrowell} 
is important, but the color assigned to any specific
particle is not.

\subsection{Comparison with schematic model}
 
We now generalize the simple model of Sec.~\ref{subsec:schem}
to include spatial correlations.
Following~\cite{Ediger_blocks}, we assume
that particles diffuse independently, but
with diffusion constants that are correlated 
over large spatial regions, each containing $n_\mathrm{c} \gg 1$ 
particles.  Since particles
diffuse independently in any given run of the dynamics,
it follows that
$\langle f_i(t) f_j(t) \riso=
\langle f_i(t) \riso\langle f_j(t)\riso$ for $j\neq i$, and hence:
\ba
\delta_{4,f}(t) &=& n_\mathrm{c} [F(2t) - F^2(t)] \nonumber \\
\chi^\mathrm{iso}_{4,f}(t) &=& (1/2) [ 1+F(4t)-2F(2t)].
\ea
In this simple model, 
both $\delta_{4,f}(t)$ and $\chi_{4,f}(t)$ scale with $n_c$,
while the isoconfigurational susceptibility 
$\chi^\mathrm{iso}_{4,f}(t)$ is not sensitive to spatial correlations
of the mobility and remains ${\cal O}(1)$. 
Thus, $\chi_{4,f}(t) \approx \delta_{4,f}(t) \gg  
\chi^\mathrm{iso}_{4,f}(t)$, and hence $R_{4,f} \simeq 1$.  
This model shows that $R_{4,f}(t)$ becomes large if 
the lengthscale for dynamic heterogeneity 
is primarily structural in nature. 
The results of Fig.~\ref{ratiolj} therefore 
indicate that this is the case for the LJ system. 

\subsection{Comparison with a kinetically constrained model}

We end with a brief discussion of
the 1-FA model,
which gives a useful theoretical insight into
the quantities discussed above. 
As recalled in Sec.~\ref{models}, 
the model describes a dynamically heterogeneous material in which
a few mobile excitations diffuse
through an immobile background. Our simulations indicate that
the ratios $R_P(t)$ and $R_{4,P}(t)$ have limiting forms at
low $T$, which depend on dimensionality, $d$.
In $1d$, the structure is very strongly
correlated with the dynamics both locally and globally:
$[R_P(\tau),R_{4,P}(\tau)] \approx [0.5,0.7]$.
However, in $3d$, the single
site ratio vanishes, $R_P(\tau)\approx0$, while the global
ratio is quite large, $R_{4,P}(\tau)\approx 0.4$.
This occurs because the set of sites visited by a given
excitation in $3d$ varies enormously from run-to-run.
Using the initial positions of excitations to predict which sites 
will relax first in a given run is 
impossible.  However,
collective observables reveal that
the rate of relaxation is reproducibly enhanced in
regions with relatively many excitations (see
also~\cite{hedges}).

This decoupling between the local and global ratios 
illustrates a situation in which the 
relation between dynamics and structure is
only statistically significant 
at large length scales.
Interestingly, this result is somewhat similar to that 
shown in Fig.~\ref{ratiolj} for the LJ system, 
although the microscopic mechanisms at work are presumably different.
Moreover, preliminary studies indicate that the strong length scale
dependence of predictability found in the 1-FA model is much less pronounced
in models where kinetic constraints are stronger.  In these 
other models, it
appears that the dynamics on all length scales is strongly constrained by
the initial structure.

\section{Outlook}
\label{conclusion}

We have investigated the degree to which structural
fluctuations in glass-forming liquids influence their
dynamical fluctuations.
We defined $R_{4,f}$ and $R_{f}$ which are quantitative
measures of this influence, on long and short length scales respectively:
they
differ by nearly an order of magnitude in the LJ system
at the lowest temperature considered.
Thus, the influence of structure on dynamics 
is much stronger on long length scales than on short ones. 
This unexpected finding constitutes our main result.

Our work does not reveal which structural features
are responsible for dynamic heterogeneity, but they
do show that the search for such an observable should be undertaken
at a coarse-grained level.  
This is consistent with 
recent studies \cite{hedges,coslovich,reichman,modes}.

As a first step towards  identifying a suitable
coarse-grained structural quantity, we exploit 
the fact~\cite{Berthier_science_long,heuer,poole} 
that potential energy is correlated with dynamical 
heterogeneities (although correlations of the energy 
remain short-ranged).  We compare fluctuations of 
the coarse-grained energy field, $\overline{e}_i(\notsigma)$, with those
of the propensity. 
By coarse-graining on a length scale $\notsigma=2$,
so that $\notsigma\simeq\xi^\mathrm{prop}$, 
we average away local fluctuations.  We obtain a field that
varies in space on a similar length scale to that of the 
dynamical propensity, and which reflects the
average energy of different regions of the system.
Interestingly, we find that regions with 
small energy are correlated with regions of low propensity
for motion, as shown in Fig.~\ref{4fig}e.  While a more
quantitative analysis is necessary before drawing firm conclusions,
this correlation between energy and propensity
is consistent with the strong local 
correlations between energy and dynamics 
demonstrated in~\cite{Berthier_science_long,heuer}. 
As an alternative to the energy, another promising route
to a connection
between structure and dynamics is provided by the presence 
of extended modes characterizing the vibrational spectrum of 
amorphous materials~\cite{modes,reichman}, and it would 
also be interesting to study connection between 
propensity and the locally ordered regions discussed in 
Ref.~\cite{pastore}.

In any case, identifying the non-local structural features that
are associated with mobile or immobile regions of glass-formers
remains a central challenge.

\acknowledgments

We thank D. Chandler, J.P. Garrahan, 
P. Harrowell, L. Hedges, and D. Reichman for discussions.
RLJ was funded by NSF grant CHE-0543158
and LB by the Joint Theory Institute at the Argonne
National Laboratory and the University of Chicago.


\begin{thebibliography}{99}

\bibitem{SRN} M.D. Ediger, C.A. Angell, and S.R. Nagel, 
J. Phys. Chem. {\bf 100}, 13200 (1996).

\bibitem{DH}
M.~D.~Ediger, Annu. Rev. Phys.  Chem. {\bf 51}, 99 (2000).

\bibitem{Kob1}  
W. Kob, C. Donati, S.J. Plimpton, P.H. Poole, and S.C. Glotzer,
Phys. Rev. Lett. {\bf 79}, 2827 (1997).

\bibitem{Ediger_blocks}
M.T.~Cicerone, P.A.~Wagner, and M.D.~Ediger,
J. Phys. Chem. B {\bf 101}, 8727 (1997). 

\bibitem{GC}
J.~P.~Garrahan and D.~Chandler, Phys. Rev. Lett. {\bf 89}, 035704
(2002). 

\bibitem{Whitelam04}
S. Whitelam, L. Berthier and J.P. Garrahan, Phys. Rev. Lett. {\bf92}, 
185705 (2004).

\bibitem{heuer}  
B. Doliwa and A. Heuer, Phys. Rev. E {\bf 67}, 031506 (2003);
J. Qian, R. Hentschke, and A. Heuer, J. Chem. Phys. {\bf 111}, 10177 (1999).

\bibitem{chi4_franz}
S. Franz, C. Donati, G. Parisi, and S.C. Glotzer,
Philos. Mag. B {\bf 79}, 1827 (1999).

\bibitem{chi4_ton}
C.~Toninelli, M. Wyart, L. Berthier, G. Biroli, and J.-P. Bouchaud, 
Phys. Rev. E {\bf 71}, 041505 (2005).

\bibitem{Berthier_science_long}
L. Berthier, G. Biroli, J.-P. Bouchaud, 
L. Cipelletti, D. El Masri, D. L'Hote, F. Ladieu, and M. Pierno, 
Science {\bf310}, 1797 (2005);
L. Berthier, G. Biroli, J.-P. Bouchaud, W. Kob, K. Miyazaki, and 
D.R. Reichman, J. Chem. Phys. {\bf126}, 184503 (2007); 
\emph{ibid}, {\bf126}, 184504 (2007).

\bibitem{pinaki} P. Chaudhuri, L. Berthier, and W. Kob,
Phys. Rev. Lett. {\bf 99}, 060604 (2007).

\bibitem{FA84} G.H. Fredrickson and H.C. Andersen,
Phys. Rev. Lett. {\bf 53}, 1244 (1984).

\bibitem{frustration_review}
G. Tarjus, S. Kivelson, Z. Nussinov, and P. Viot,
J. Phys.: Condens. Matter {\bf 17}, R1143 (2005).

\bibitem{AG}
G.~Adam and J.~H.~Gibbs, J. Chem. Phys. {\bf 43}, 139 (1958).

\bibitem{KTW}
T.R. Kirkpatrick and D. Thirumalai, Phys. Rev. Lett.
{\bf 58}, 2091 (1987); T.R. Kirkpatrick and P. Wolynes, Phys. Rev. B
{\bf 36}, 8552 (1987); T.R. Kirkpatrick, D. Thirumalai and P. Wolynes,
Phys. Rev. A {\bf 40}, 1045 (1987).

\bibitem{MCT}
W. G\"{o}tze and L. Sj\"{o}gren, Rep. Prog. Phys. {\bf 55}, 55 (1992);

\bibitem{modes}
C. Brito and M. Wyart, Europhys. Lett. {\bf 76}, 149 (2006);
C. Brito and M. Wyart, J. Stat. Mech. (2007) L08003.
\bibitem{reichman}
A. Widmer-Cooper, P. Harrowell, H. Perry and 
D.R. Reichman (unpublished).

\bibitem{Stillinger-Weber}
F. Stillinger and T. Weber, Science {\bf 225}, 983 (1984).

\bibitem{Harrowell} 
A. Widmer-Cooper, P. Harrowell, and H. Fynewever, 
Phys. Rev. Lett. {\bf 93}, 135701 (2004);
A.~Widmer-Cooper and P.~Harrowell, J. Phys.: Cond. Matt. {\bf17},
S4025 (2005); Phys. Rev. Lett. {\bf 96}, 185701 (2006).

\bibitem{Harro07}
A. Widmer-Cooper and P. Harrowell, J. Chem. Phys. {\bf126}, 154503 (2007)

\bibitem{poole}
G.S.~Matharoo, M.S.H.~Razul and P.H.~Poole, Phys. Rev. 
E {\bf74}, 050502 (2006).

\bibitem{appig}
G.A.~Appignanesi, J.A.R.~Fris and M.A.~Frechero, Phys. Rev. Lett. {\bf 96},
237803 (2006);
J.A. Rodriguez Fris, L.M. Alarc\'on, and G.A. Appignanesi,
Phys. Rev. E {\bf 76}, 011502 (2007).

\bibitem{coslovich}
D.~Coslovich and G.~Pastore, Europhys. Lett. {\bf75}, 784 (2006).

\bibitem{hedges}
L.O.~Hedges and J.P.~Garrahan,  J. Phys.: Cond. Matt. {\bf19}, 205124 (2007).

\bibitem{frechero}
M.A.~Frechero, L.M. Alarc\'on, E.P. Schulz, and G.A. Appignanesi,
Phys. Rev. E {\bf 75}, 011502 (2007).

\bibitem{kennet}
M.~T.~Downton and M.~P.~Kennett, arXiv:0704.1497.

\bibitem{KA} W. Kob and H.~C. Andersen,
Phys. Rev. Lett. {\bf 73}, 1376 (1994).

\bibitem{BKS}
B.W.H.~van Beest, G.J.~Kramer, and R.A.~van Santen, 
Phys. Rev. Lett. {\bf 64}, 1955 (1990). 

\bibitem{Berthier_MC_LJ} L. Berthier and W. Kob,
J. Phys.: Condens. Matter {\bf 19}, 205130 (2007).

\bibitem{Berthier_MC_BKS}
L. Berthier, Phys. Rev. E {\bf 76}, 011507 (2007). 

\bibitem{Jung04}
Y.~Jung, J.~P.~Garrahan and D.~Chandler, Phys. Rev. E {\bf 69},
061205 (2004).

\bibitem{potato}
P. Harrowell, private communication.

\bibitem{foot} 
We obtain $\xi^{\rm prop}$
by collapsing $S_4^{\rm prop}$ 
onto a master curve, $S_4^{\rm prop}(q,t)/
S_4^{\rm prop}(0,t) = f(q \xi^{\rm prop})$, with 
$f(x) = 1/(1+x^2+x^5)$ suggested 
by the numerical data~\cite{Whitelam04,chi4_ton}. 

\bibitem{pastore}
D. Coslovich and G. Pastore,  
preprint arXiv:0705.0626 and arXiv:0705.0629.

\end{thebibliography}
\end{document}